\documentclass[fleqn,usenatbib]{mnras}

\usepackage{newtxtext,newtxmath}

\usepackage[T1]{fontenc}

\DeclareRobustCommand{\VAN}[3]{#2}
\let\VANthebibliography\thebibliography
\def\thebibliography{\DeclareRobustCommand{\VAN}[3]{##3}\VANthebibliography}

\usepackage{graphicx}	
\usepackage{amsmath}	

\title[Flare-Driven Atmospheric Variability]{Can the Long-Term Impact of Stellar M-Dwarf Flares Alter the Spectral Features of a Giant Gaseous Exoplanet?}

\author[A. J. Louca et al.]{
Amy J. Louca,$^{1}$
Shang-Min Tsai$^{2}$,
and Yamila Miguel$^{1,3}\thanks{E-mail: ymiguel@strw.leidenuniv.nl}$
\\
$^{1}$SRON, Netherlands Institute for Space Research, Niels Bohrweg 4, NL-2333 CA, Leiden, The Netherlands\\
$^{2}$Institute of Astronomy and Astrophysics, Academia Sinica, No.1, Sec. 4, Roosevelt Rd, Taipei 106319, Taiwan, R.O.C\\
$^{3}$Leiden Observatory, Leiden University, Einsteinweg 55, 2333 CA Leiden, The Netherlands\\
}

\date{Accepted XXX. Received YYY; in original form ZZZ}

\pubyear{2026}

\begin{document}
\label{firstpage}
\pagerange{\pageref{firstpage}--\pageref{lastpage}}
\maketitle

\begin{abstract}
In this work, we model the long-term impact of recurrent stellar flares on the atmospheres of metal-rich gaseous exoplanets. Using synthetic flare spectra from a fiducial flare model integrated with a photochemical kinetics code, we track the changes in atmospheric composition with time. We further analyze the spectral variability by feeding these abundance profiles into a radiative transfer code at various time steps. Our simulations showed variability and persistent changes in key atmospheric species, such as CH$_4$, CO$_2$, and SO$_2$, when compared to their quiescent state. Extreme flare events cause rapid depletion of molecules in the upper atmosphere and a temporary disappearance of spectral features, especially the SO$_2$ feature at 7-8 microns, which shifted by about 75 ppm. Many species did not fully return to their quiescent state after flares, resulting in lasting changes in abundance, especially for SO$_2$ and CO$_2$, key species when inferring the atmospheric metallicty. We also explored the cumulative effects of recurrent flares, showing that species like H$_2$O and CH$_4$ followed a decreasing abundance trend, with half-lives of around 28 to 31 years. These results indicate that flare activity plays a significant role in shaping both the short- and long-term atmospheric composition and spectral features of giant gaseous exoplanets orbiting M-dwarf stars, underscoring the need to account for stellar activity when characterising such atmospheres. Our findings also highlight that the atmosphere of the modelled planet is not static, suggesting that a probabilistic approach to atmospheric abundances may be more appropriate than static retrievals, particularly for gaseous planets orbiting active stars.
\end{abstract}




\section{Introduction}

Exoplanet atmospheres are heavily affected by environmental conditions, such as stellar irradiation and winds. This influence can be intense and occur in short, high-energy bursts, especially around M-dwarf stars (e.g., \citeauthor{shkolnik2014} \citeyear{shkolnik2014}; \citeauthor{Schneider2018} \citeyear{Schneider2018}; \citeauthor{Loyd2018a} \citeyear{Loyd2018a}). 
M-dwarfs are known for producing frequent, high-energy outbursts, such as flares, that expose orbiting planets to extreme ultraviolet (XUV) and X-ray radiation. The XUV spectrum of these stars are, therefore, more energetic and follows a different shape than what we expect to see compared to stars like our Sun. The MUSCLES collaboration (\citeauthor{France2016} \citeyear{France2016}; \citeauthor{Youngblood2016} \citeyear{Youngblood2016}; \citeyear{Youngblood2017}; \citeauthor{Loyd2016} \citeyear{Loyd2016}; \citeyear{Loyd2018a}; \citeauthor{Brown2023} \citeyear{Brown2023}) has mapped the spectra of these stars in detail by observing various M- and K-dwarf stars across optical to X-ray wavelengths. These stars showed signs of flaring activity during some of the observations, giving more insights into the activity of low-mass stars (e.g., \citeauthor{Loyd2018a} \citeyear{Loyd2018a}).  

Given that most exoplanets discovered to date orbit M-dwarfs, the impact of these environmental factors is especially relevant. Although smaller and cooler than stars like the Sun, M-dwarfs are some of the most magnetically active in the galaxy, producing stellar activity that can persist for billions of years. This intense radiation can drive atmospheric evaporation, alter chemical compositions, and strip away gases from the upper atmosphere (see e.g., \citeauthor{Vidal-Madjar2003} \citeyear{Vidal-Madjar2003}; \citeauthor{Segura2010} \citeyear{Segura2010}; \citeauthor{Ehrenreich2015} \citeyear{Ehrenreich2015};  \citeauthor{Venot2016} \citeyear{Venot2016}; \citeauthor{Chen2021} \citeyear{Chen2021}; \citeauthor{Konings2022} \citeyear{Konings2022}; \citeauthor{Louca2023} \citeyear{Louca2023}; \citeauthor{Nicholls2023} \citeyear{Nicholls2023}). Initially, most planets orbiting M-dwarfs were believed to be mainly small, Earth-sized worlds. However, recent discoveries have revealed that larger, more massive planets, including gas giants, exist around these relatively small stars (\citeauthor{Jordan2022} \citeyear{Jordan2022}; \citeauthor{Stefansson2023} \citeyear{Stefansson2023}; \citeauthor{Kanodia2024} \citeyear{Kanodia2024}). These consequences extend
to flaring specifically: flare-driven mass loss can be significant for planets
orbiting active M-dwarfs and scales strongly with orbital separation
(\citeauthor{Amaral2025} \citeyear{Amaral2025}), while flares can also
perturb the thermal structure, circulation, and chemistry of planetary
atmospheres through coupled three-dimensional feedbacks
(\citeauthor{Chen2025} \citeyear{Chen2025}). These findings challenge earlier assumptions and suggest that the planetary systems of M-dwarfs are more diverse than previously thought. This has made such systems increasingly valuable for in-depth study, as they could offer new insights into atmospheric evolution in environments different from those of larger stars (e.g. JWST-GEMS GO program 3171).

Various theoretical work has already been done on these effects. \cite{Louca2023} (hereafter L23) showed that recurrent flaring of $\sim$11 days may result in accumulating change within hydrogen-dominated exoplanet atmospheres. However, they argued that a more thorough investigation with longer integration times is needed to test whether compositional changes accumulate over time due to recurrent flaring. This study limited the chemical network to molecular species containing nitrogen (N), carbon (C), oxygen (O), and hydrogen (H), therefore excluding photochemical-sensitive species such as SO$_2$, OCS, and CS$_2$, containing sulfur (S). It was shown recently, however, that these photochemical-sensitive species, like SO$_2$, are present in the atmospheres of gaseous exoplanets (e.g., \citeauthor{Tsai2023} \citeyear{Tsai2023}; \citeauthor{Holmberg2024} \citeyear{Holmberg2024}; \citeauthor{Dyrek2024} \citeyear{Dyrek2024}; \citeauthor{Beatty2024} \citeyear{Beatty2024}; \citeauthor{Gressier2025} \citeyear{Gressier2025}). Including such a chemical network is, therefore, crucial in understanding how it might change our observations.

Typically, the compositions of exoplanet atmospheres have been considered relatively static, given the long planetary lifetimes. However, this perspective overlooks the potential for significant short-term variability driven by stellar activity. Despite extensive theoretical work to model these effects, observational evidence of variable atmospheres is still lacking. With the arrival of the James Webb Space Telescope (JWST) and its high-precision observational capabilities, we now face a critical question: \textit{Can we fully trust the planetary spectra at a given moment in time, or should we account for short-term variability due to stellar activity when interpreting these results?}

This work focuses on the short- and long-term variability of high-metallicity atmospheres to see how variable these photo-chemical species are due to recurrent flaring events. We extend the study of \citet{Louca2023} and examine the longer-duration effect of recurrent flaring to reveal the (non-)accumulative effect. We conduct simulations of the atmosphere of a hypothetical hot Jupiter planet orbiting an M-dwarf star over an integration period of $\sim$3 years. Additionally, we include sulfur chemistry to investigate photochemical-sensitive species such as SO$_2$. We perform a statistical analysis to estimate the probability of observing variability within exoplanet atmospheres with the currently most advanced observing facility for characterising exoplanet atmospheres, JWST. 

\section{Methods}
\label{sec:MethodsP4}

Given the computational cost of time-dependent photochemical simulations over multi-year timescales, each model requires several months of computation. We therefore restrict this first study to a single representative planetary case, allowing us to investigate the underlying physical processes rather than perform a statistical exploration of parameter space.

\subsection{Planetary system}

In this study, we conduct simulations of a Jupiter-mass exoplanet orbiting a low-mass star with a radius of 0.36 $R_{\odot}$ and an effective temperature of $T=$ 3350K. The planet is positioned at a distance of 0.01 AU from the host star, giving the synthetic planet an equilibrium temperature of $T = 1370$ K. To maintain simplicity and focus on fundamental orbital dynamics, we incorporate low eccentricity into the system. This approach allows us to explore the thermal and dynamical characteristics of the exoplanet-star system under controlled conditions, providing valuable insights into its atmospheric behavior. Finally, we assume the gas giant ($M = 1 M_J$; $R = 1 R_J$) to be hydrogen-dominated with a metallicity of 10xsolar-like. This is to ensure SO$_2$ is abundant enough to be observed (\citeauthor{Polman2023} \citeyear{Polman2023}; \citeauthor{Tsai2023} \citeyear{Tsai2023}; \citeauthor{Powell2024} \citeyear{Powell2024}).

\subsection{Stellar spectra and lightcurves}

\subsubsection{Quiescent spectrum}
Quiescent stellar spectra are obtained from the Measurements of the Ultraviolet Spectra Characteristics of Low-mass Exoplanetary Systems (MUSCLES) collaboration (\citeauthor{France2016} \citeyear{France2016}; \citeauthor{Youngblood2016} \citeyear{Youngblood2016}; \citeyear{Youngblood2017}; \citeauthor{Loyd2016} \citeyear{Loyd2016}; \citeyear{Loyd2018a}; \citeauthor{Brown2023} \citeyear{Brown2023}). The MUSCLES survey comprises observations of M- and K-dwarf exoplanet host stars across optical, ultraviolet (UV), and X-ray wavelengths, collected via instruments including the Hubble Space Telescope (FUV - blue visible), the Chandra Xray Observatory, and the XMM-Newton. An openly accessible database is provided, containing panchromatic spectral energy distributions (hereafter SEDs) spanning $5\cdot10^{-4}$ to 5.5 $\mu$m. This study uses the SED of GJ 876 as observed and delivered by the MUSCLES collaboration.  

\begin{figure*}
    \centering
    \includegraphics[width=0.8\textwidth]{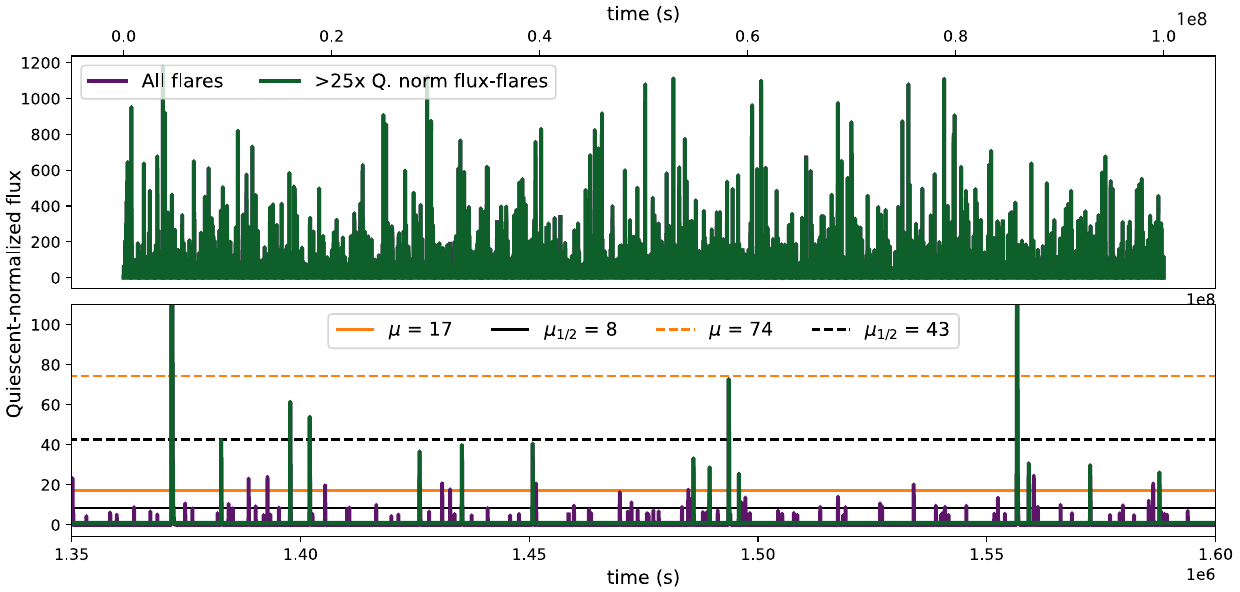}
    \caption{The lightcurve, normalized to the quiescent level, is generated using the fiducial flare model. The top panel displays the full lightcurve, including all detected flares. In contrast, the bottom panel offers a closer view of the lightcurve between 1.35e6 and 1.6e6 seconds, highlighting the difference between cases with many flares (in purple) and fewer flares (in green). The solid black and orange lines represent the median and mean flux levels, respectively, for all flares. Meanwhile, the dashed black and orange lines indicate the median and mean flux levels when only flares exceeding 25 times the quiescent-normalized flux are considered.}
    \label{fig:lightcurve}
\end{figure*}

\subsubsection{Synthetic flare spectra}

To assess the prolonged impact of stochastic flaring from the host star, we utilize the fiducial flare program developed by \cite{Loyd2018a}. This program enables us to extend our analysis beyond the temporal constraints of existing UV observations. The fiducial flare spectrum is constructed based on energy allocations derived from empirical flare measurements. These spectra are calibrated to match the Si IV doublet of observed spectra from the MUSCLES-spectra dataset. To integrate the flaring events into our synthetic spectra, we incorporate quiescent spectra, ensuring that the energy levels of included flares consistently surpass or equal those of the quiescent state. We adopt a single fiducial flare spectrum that is scaled in intensity by a time-dependent factor, such that the spectral shape remains fixed while only the amplitude varies throughout each flare event.
\begin{equation}
    F_{f}(t) = F_q \cdot (a(t) +1)
\end{equation}
Where $F_f$ is the fiducial flare spectrum, $F_q$ is the quiescent spectrum and $a(t)$ is the scaling factor, as described over time by figure \ref{fig:lightcurve}.
The final lightcurve is generated by creating 100 individual lightcurves, each with a duration of $10^{6}$ seconds and a bin width of 10 seconds. This is to maintain the high-resolution temporal profile of each fiducial flare event. All individual lightcurves are then stitched together. The final lightcurve is presented in Figure \ref{fig:lightcurve}. The top panel shows the lightcurve with all flares included, alongside the lightcurve where only flares exceeding 25 times the quiescent flux are considered. The difference between these two cases is subtle in this panel. The bottom panel provides a zoomed-in view from 1.35$\cdot10^{6}$ to 1.6$\cdot10^{6}$ seconds. Excluding smaller flares results in a significant increase in both the mean ($\mu$) and median ($\mu_{1/2}$) flux values, with $\mu$ rising from 17 to 74, and $\mu_{1/2}$ from 8 to 43.  L23 demonstrated that including smaller flares is essential for more accurate assessments of flare impacts but also highlighted the computational challenges of including the full lightcurve into the chemical kinetics code. To address this, we will consider both cases in our simulations to evaluate the long-term effects of stellar flaring under different scenarios. Flare energies and equivalent durations are defined over the Si IV doublet (1393.76 \AA and 1402.77 \AA), following the methodology of L23 (see section 2.2.2) and \citeauthor{Loyd2018b} \citeyear{Loyd2018b}. The full SED of GJ 876 that has been used can be found in L23 figure 1.

\subsection{Atmospheric model}

\subsubsection{Chemical kinetics}

The chemical evolution is modeled using the open-source chemical kinetics code \texttt{VULCAN} (\citeauthor{vulcan2017} \citeyear{vulcan2017}; \citeyear{Tsai2021}) which also includes photo-chemistry. The model contains a chemical network involving H, N, C, O, and S, encompassing 1089 forward, backward, and photochemical reactions. A metallicity of 10 times the solar elemental abundance is assumed to ensure that species like SO$_2$ are present and detectable in the atmosphere (\citeauthor{Polman2023} \citeyear{Polman2023}). Another variable to which SO$_2$ is sensitive is the C/O ratio. For sub-solar C/O ratios, more oxygen becomes available to form SO$_2$. In contrast, for super-solar values, the oxygen-to-carbon content becomes unbalanced and is unable to create significant amounts of SO$_2$ (\citeauthor{Polman2023} \citeyear{Polman2023}). In this study, we sub-solar value of C/O $\approx$ 0.46.

\begin{figure}
    \centering
    \includegraphics[width=1.\linewidth]{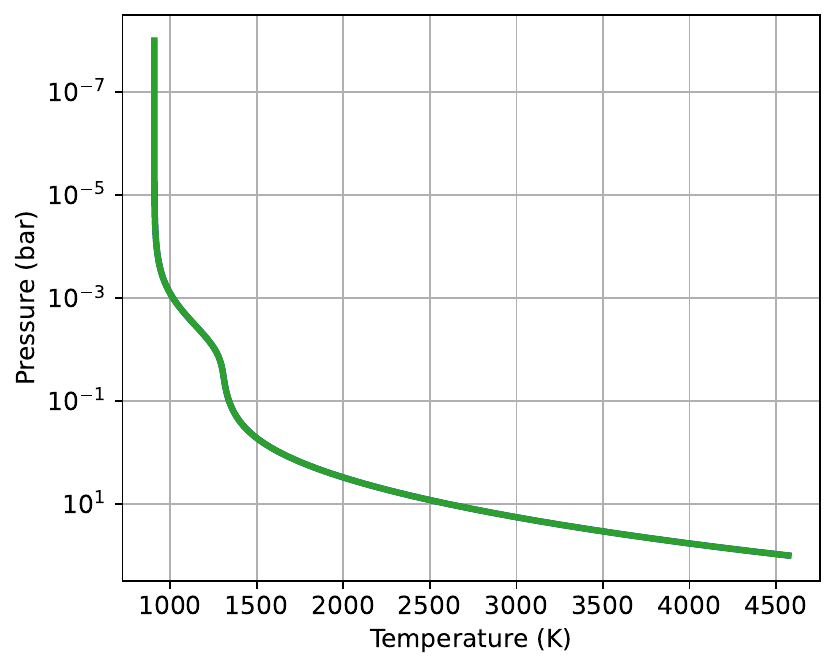}
    \caption{The temperature-pressure profile of a warm Jupiter orbiting an M-dwarf at a distance of $a = 0.01$ AU. }
    \label{fig:TP}
\end{figure}

The simulations include an analytically derived temperature-pressure profile (TP-profile) from \cite{Heng14} to model the temperature-pressure distribution within the exoplanet's atmosphere. For this, we assume an internal temperature of $T_\mathrm{int} =$ 300 K, an equilibrium temperature of $T_{\mathrm{eq}} = $ 1370 K, and default opacity parameters of $\kappa_{\mathrm{L}} = 0.1$, $\kappa_{\mathrm{S}} = 0.02$, $\beta_{\mathrm{L}} = 1$, and $\beta_{\mathrm{S}} = 1$. The final TP-profile is shown in figure \ref{fig:TP}.

Both molecular and eddy diffusion are considered for vertical mixing. We use an eddy diffusion constant of $K_{zz} = 10^{10}$ cm$^{2}$ s$^{-1}$, which also has previously been adopted by other studies (e.g., \citeauthor{Moses2012} \citeyear{Moses2012}; \citeauthor{Parmentier2013} \citeyear{Parmentier2013}; \citeauthor{Miguel2014} \citeyear{Miguel2014}). 

Finally, to ensure similarities with L23, we also assume an open upper boundary condition with an upward flux term due to the escape of light particles. This study assumes that this escape is limited at the homopause due to diffusion, as defined in \citet{Tsai2021}. 

\subsubsection{Radiative transfer}

\begin{table*}
    \centering
    \begin{tabular}{c|c|c|c}
        \textbf{Molecule} & \textbf{Temperature range (K)} & \textbf{Pressure range (bar)} & \textbf{Line lists \texttt{pRT}}  \\
        \hline
        \hline
        H$_2$O & 50 - 2900 & $10^{-8}$ - $10^3$ & POKAZATEL$^1$ \\
        CH$_4$ & 50 - 2900 & $10^{-8}$ - $10^3$ & YT34to10$^2$ \\
        SO$_2$ & 50 - 2900 & $10^{-8}$ - $10^3$ & ExoAmes (v2)$^3$\\
        CO & 50 - 2900 & $10^{-8}$ - $10^3$ & HITEMP$^{5}$ \\
        H$_2$S & 50 - 2900 & $10^{-8}$ - $10^3$ &  AYT2$^4$\\
        CO$_2$ & 50 - 2900 & $10^{-8}$ - $10^3$ & UCL-4000$^{6}$\\
        NH$_3$ & 50 - 2900 & $10^{-8}$ - $10^3$ &  CoYuTe$^9$ \\
        CS$_2$ & 50 -2900 & $10^{-8}$ - $10^3$ & HITRAN2020$^{7}$ \\
        OCS & 50 - 2000 & $10^{-5}$ - $10^2$ & OYT8$^8$ \\
        H$_2$ & 50 -2900 & $10^{-8}$ - $10^3$ & HITRAN2012$^{10}$ 
    \end{tabular}
    \caption{ The line lists used for each opacity species included in radiative transfer. [1] \citet{Polyansky2018}; [2] \citet{Yurchenko2014}; \citet{Yurchenko2017} [3] \citet{Underwood2016}; [4] \citet{Azzam2016}; \citet{CHUBB2018178}; [5] \citet{ROTHMAN20102139}; [6] \citet{Yurchenko2020}; [7] \citet{hitran2020}; [8] \citet{owens2024}; [9] \citet{Coles2019}; [10] \citet{rothman2013}}
    \label{tab:linelists}
\end{table*}

To simulate transmission spectra, we make use of the open-source radiative transfer code petitRADTRANS (\citeauthor{Molliere2019} \citeyear{Molliere2019}; \citeyear{Molliere2020}; \citeauthor{prt2022} \citeyear{prt2022}). We include opacity sources from H$_2$O, CH$_4$, SO$_2$, CO$_2$, CO, NH$_3$, H$_2$S, OCS, CS$_2$, and H$_2$ as listed in table \ref{tab:linelists}. We also include collisional induced absorption (CIA) from H$_2$-H$_2$ and H$_2$-He. The resolution element for all opacity sources is set to R = 1000. 
We ignore any cloud decks and focus solely on the change in molecular features in the transmission spectrum. The transmission spectrum is determined for several timesteps within the simulation. To begin with, we look at the spectrum when the atmosphere is in a quiescent state before any flare events occur. We then focus on the change in transmission spectrum during a large flare event to see how photochemical-sensitive molecular features change over time. Finally, we also want to see how variable the transmission spectrum is by calculating the transmission spectrum 13000 times in a timespan of $10^7$ seconds. 

\section{Results}

\label{sec:ResultsP4}

\subsection{Atmospheric composition}

\subsubsection{Quiescent state}

\begin{figure}
    \centering
    \includegraphics[width=1.0\linewidth]{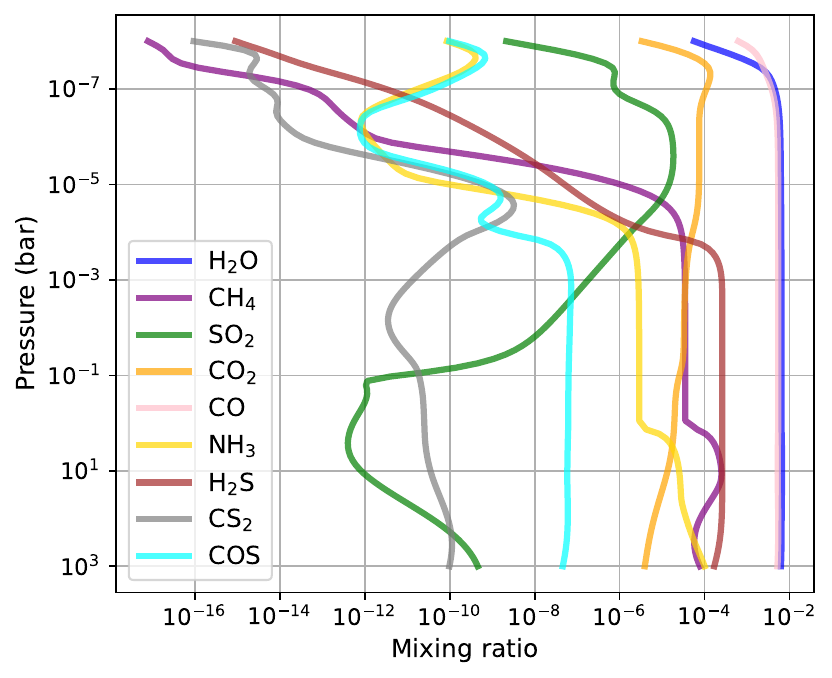}
    \caption{The mixing ratios as a function of the pressure of various molecules in a quiescent state (i.e., no flare events have occurred).}
    \label{fig:quiescent_composition}
\end{figure}

Before adding the high energy stellar flares to the simulation, the atmosphere is brought into its quiescent state. Figure \ref{fig:quiescent_composition} shows the composition of the atmosphere in this quiescent state for the molecules H$_2$O, CH$_4$, SO$_2$, CO$_2$, CO, NH$_3$, H$_2$S, CS$_2$, and COS. This figure shows that the most abundant molecules are \textbf{CO} and \textbf{H$_2$O}. Both molecules have the same relative abundance throughout most of the atmosphere. In the upper atmosphere, they both break down due to photochemical dissociation. The next most abundant molecules in the upper atmosphere are \textbf{CO$_2$} and \textbf{SO$_2$}. These molecules both owe their high abundance to the high metallicity in the atmosphere. For SO$_2$, this is enhanced even more due to photochemical reactions. The most relevant reaction in creating this molecule is SO + OH $\rightarrow$ SO$_2$ + H. Due to the photodissociation of H$_2$O, more OH becomes available to create SO$_2$ (for more details see \citeauthor{Polman2023} \citeyear{Polman2023} and \citeauthor{Tsai2023}\citeyear{Tsai2023}). 

Going a bit deeper into the atmosphere ($P > 10^{-4}$ bar), the most abundant species are \textbf{CH$_4$} and \textbf{H$_2$S}. These molecules are destroyed in the upper atmosphere due to the combination of low-pressure, high-temperature, and photochemical reactions. Methane is primarily dissociated through the reactions CH$_4$ $\rightarrow$ CH$_2$ + H$_2$ and CH$_4$ $\rightarrow$ CH$_3$ + H at pressures below $10^{-5}$ bar. It is mainly dissociated at higher pressures into CH$_4$ $\rightarrow$ OH + CH$_3$. However, methane formation occurs more rapidly at these pressure levels through the reaction CH$_3$ + HCO $\rightarrow$ CO + CH$_4$, balancing out the over-excessively available CH$_3$ from photo-dissociation. 
The final photochemical-sensitive species plotted in this figure are \textbf{COS} and \textbf{CS$_2$}. COS is the most abundant of these molecules in most parts of the atmosphere. The formation of COS in the upper atmosphere starts similar to the formation of SO$_2$ - with photolysis of water due to stellar irradiation. Figure 2 of \citet{Tsai2023} shows a schematic view of COS production.

COS mainly dissociates through the photochemical reaction COS $\rightarrow$ CO + S. However, this dissociation rate is lower than the production rate of COS, and thus, it persists in the atmosphere.  

\subsubsection{Extreme flare event}
\label{sec:extreme_flare_event_chem}

\begin{figure*}
    \centering
    \includegraphics[width=0.9\linewidth]{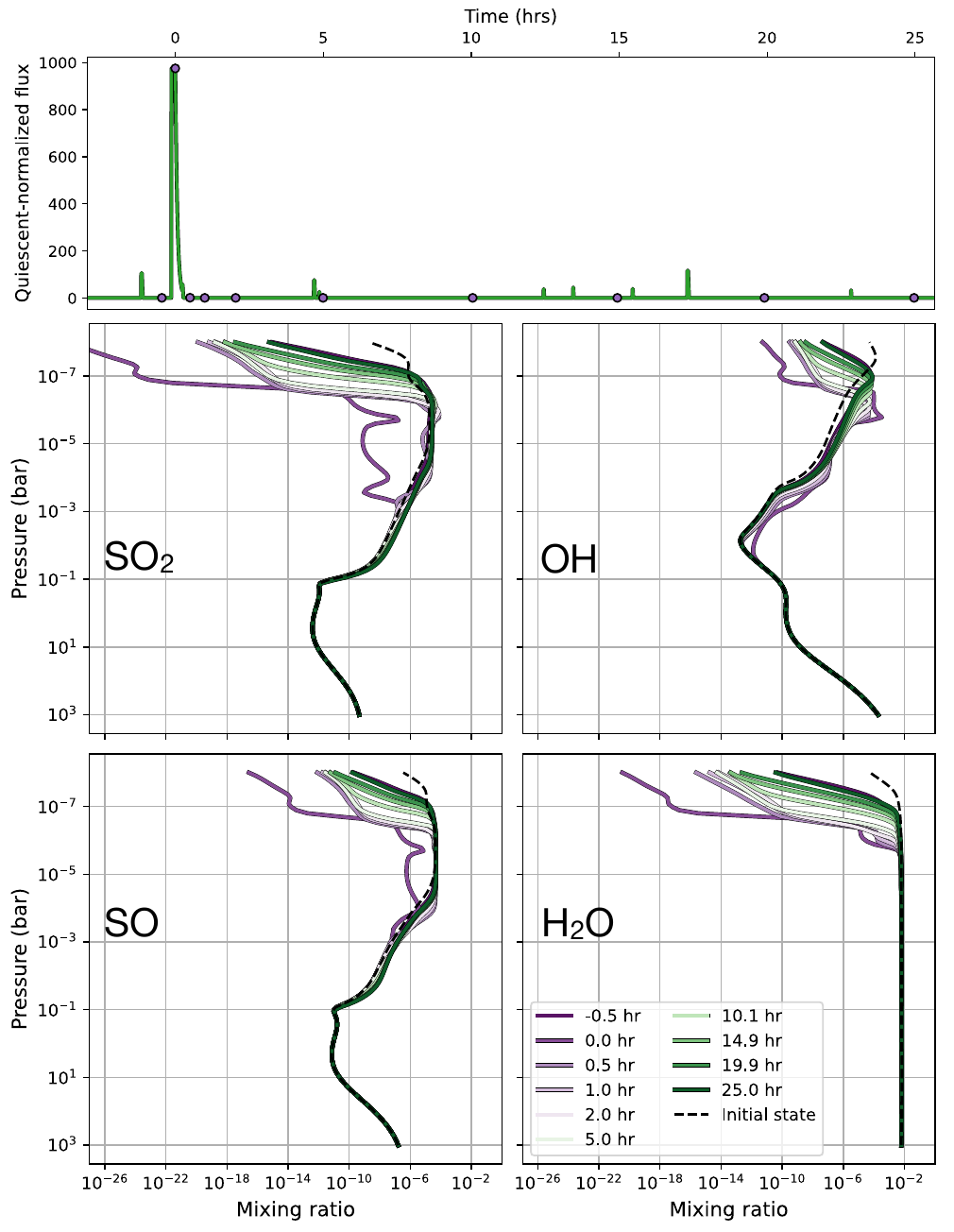}
    \caption{The change in mixing ratios of the four photochemical species SO$_2$ (center left), OH (center right), SO (lower left), and H$_2$O (lower right), during an extreme flare event. The colors indicate the time at which the snapshot of the atmosphere was taken relative to the flare event. The dashed lines represent the initial state of the atmosphere when stellar flares are not considered. The top panel represents the light curve of the extreme flare event (green line). The purple scatter points illustrate the snapshots taken from the photochemical kinetics code. The time labels are centered around the large flare event.}
    \label{fig:sulfursp_extreme_case}
\end{figure*}

\begin{figure*}
    \centering
    \includegraphics[width=0.9\linewidth]{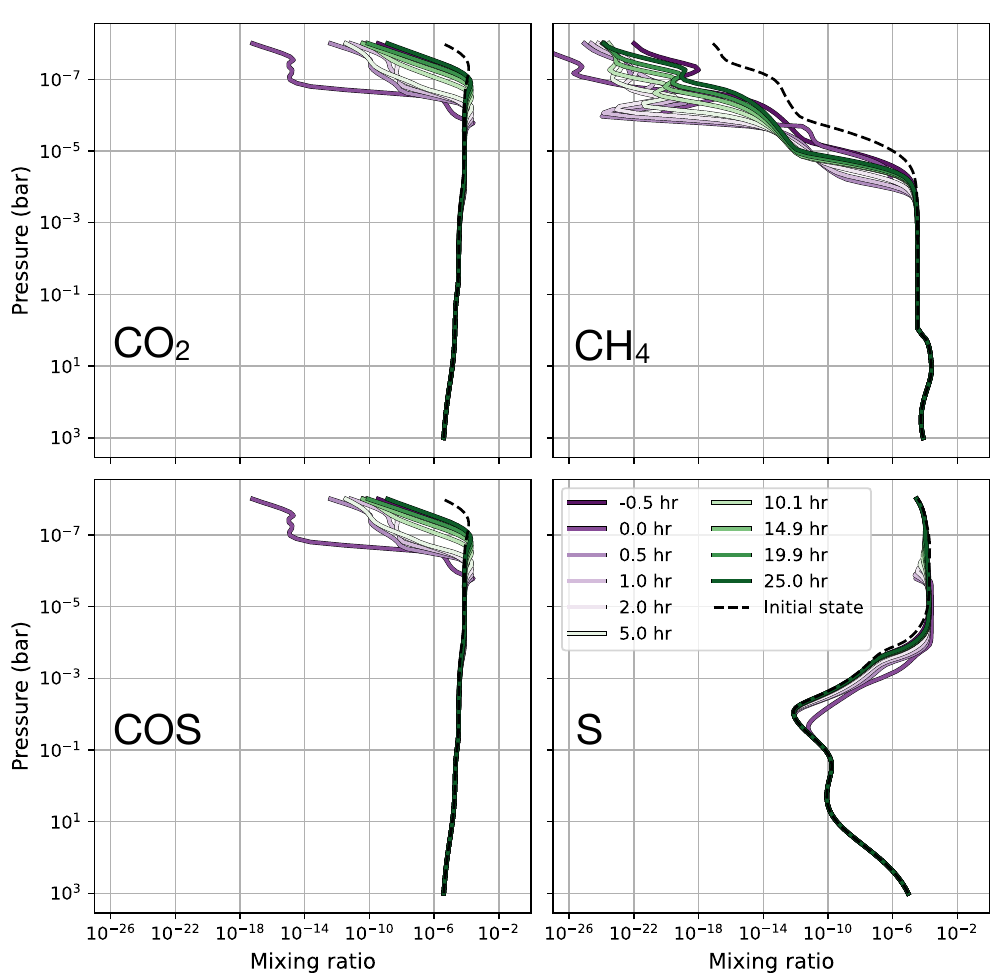}
    \caption{The change in mixing ratios of the molecules CO$_2$ (upper left), CH$_4$ (upper right), COS (lower left), and S (lower right) during an extreme flare event, similar as to figure \ref{fig:extreme_case_BIG}.}
    \label{fig:extreme_case_BIG}
\end{figure*}

Figures \ref{fig:sulfursp_extreme_case} and \ref{fig:extreme_case_BIG} illustrate the chemical evolution of the species SO$_2$, OH, SO, H$_2$O, CO$_2$, CH$_4$, COS, and S during the large flare event, as seen in the top panel of figure \ref{fig:sulfursp_extreme_case}. It can be seen in this top panel that right after the large flare event, there are still some minor flare events with a quiescent-normalized flux of $<100$. Nevertheless, the chemical evolution of each molecular species begins to gradually return to its initial state approximately 0.5 hours after the extreme flare event. This indicates that while extreme flare events significantly alter the composition of exoplanet atmospheres, these extreme changes are not permanent. 

The biggest molecular change can be found in \textbf{SO$_2$}. This molecule is photochemically produced but also subject to direct dissociation, resulting in the Gaussian-like vertical distribution \citet{Tsai2023}. SO$_2$ shows extreme photodissociation between $P= 10^{-8}$ - 10$^{-3}$ bar and is depleted in this region during the extreme flare event through: 
\begin{equation*}
 \mathrm{\ \ \ \ SO_2} \rightarrow \mathrm{SO \ } + \mathrm{\ O}
\end{equation*}
and
\begin{equation*}
 \mathrm{\ \ \ \ SO_2} \rightarrow \mathrm{S \ } + \mathrm{\ O_2}
\end{equation*}
At the same time, additional atomic S is released from H$_2$S, and OH is produced from dissociation of H$_2$O (see \ref{sec:extreme_flare_event_chem} and \ref{fig:extreme_case_BIG}). As a result, about 30 minutes after the flare, both SO and SO$_2$ reform in the intermediate atmosphere and exhibit enhanced abundances relative to the initial state, due to the extra S and OH produced during the flare. 

In parallel, this leads to an underproduction of H$_2$O due to the excess of OH is used for the formation of SO$_2$ rather than H$_2$O. 
This, in turn, gives excess oxygen, which can then be used to form \textbf{CO$_2$} through the reaction OH + CO $\rightarrow$ H + CO$_2$. Figure \ref{fig:sulfursp_extreme_case} shows an overabundance of CO$_2$ in the upper atmosphere around $10^{-7}$ bar $< P < 10^{-6}$ bar, the same region where there is a depletion of H$_2$O. At lower pressure levels, the CO$_2$ abundance shows a decrease compared to initial levels. The main dissociation reaction in this region of CO$_2$ is

\begin{equation*}
    \mathrm{\ \ \ \ CO_2} \rightarrow \mathrm{CO \ } + \mathrm{\ O}
\end{equation*}

Another photosensitive species is \textbf{CH$_4$}. Compared to SO$_2$, CH$_4$ photolysis requires a hiogher energy threshold, predominately by Lyman-$\alpha$ photons. This molecule gets dissociation in the upper atmosphere ($P < 10^{-4}$ bar). Unlike SO$_2$ and CO$_2$, the dissociation of CH$_4$ is limited, and its molecular abundance does not return to quiescent state levels. Instead, it is broken down and left in lower abundance. The main species in which methane is dissociated are CH$_3$, CH$_2$, H$_2$, and H, through photochemical reactions. 

\begin{equation*}
 \mathrm{\ \ \ \ CH_4} \rightarrow \mathrm{CH_3 \ } + \mathrm{\ H}
\end{equation*}
and
\begin{equation*}
 \mathrm{\ \ \ \ CH_4} \rightarrow \mathrm{CH_2 \ } + \mathrm{\ H_2}
\end{equation*}
 Finally, \textbf{COS} shows an overall depletion in the upper atmosphere, with some exceptions at certain pressure levels. The most interesting spike in abundance can be seen precisely during the flare event at around $P \approx 10^{-4}$ bar. This overabundance of COS shows the exact opposite effect of what happens with SO$_2$ during the flare event at this particular pressure level. This is because the decreasing production rate of SO$_2$ leaves the atmosphere with excess OH, which is used for the production of COS through the reaction
\begin{equation*}
    \mathrm{\ \ \ \ CS} + \mathrm{\ OH} \rightarrow \mathrm{COS \ } + \mathrm{\ H}
\end{equation*}
COS is broken down into CO and S at the other pressure levels through photochemistry, leaving a depleted COS atmosphere behind.

\subsubsection{Full flares comparison}
\label{sec:morefewer}
\begin{figure*}
    \centering
    \includegraphics[width=1.0\linewidth]{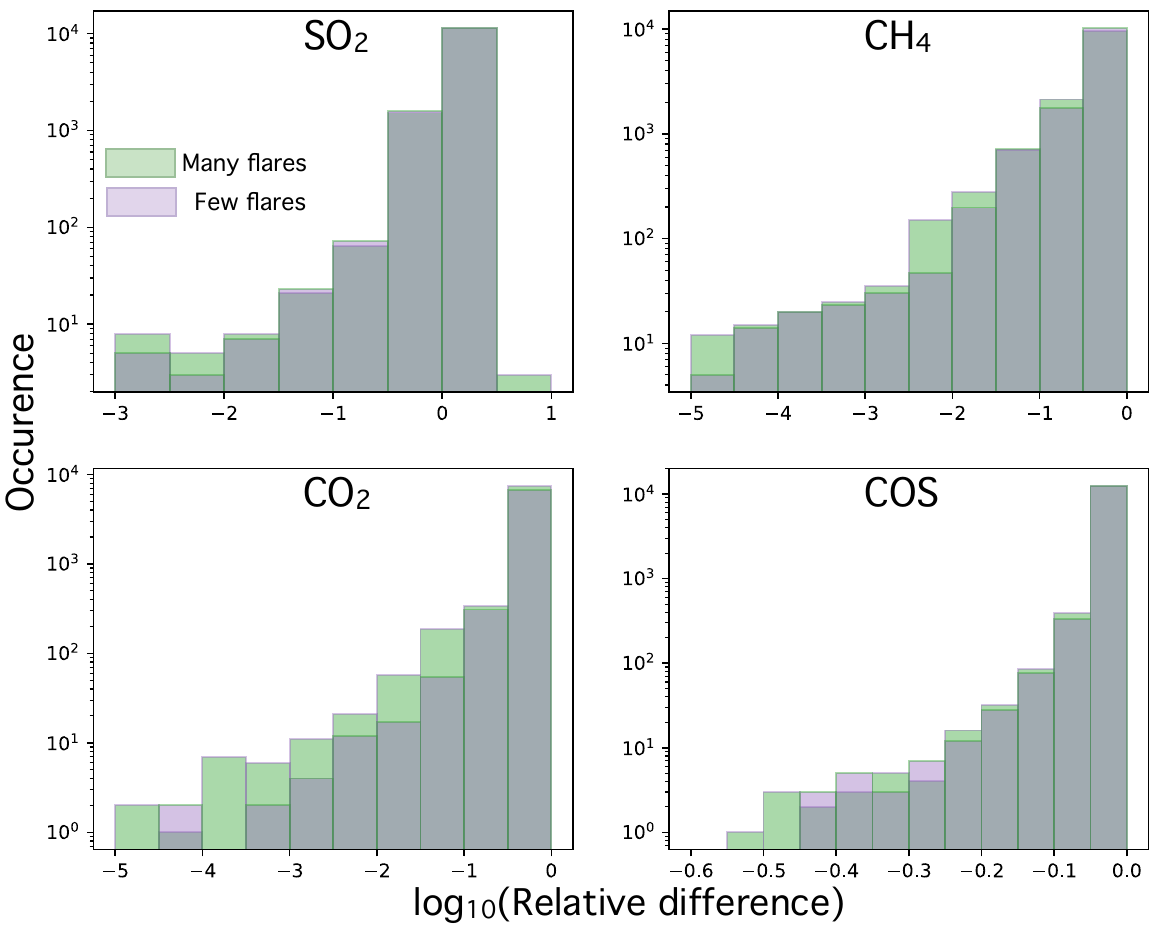}
     \caption{The logarithm of the relative difference in column number density in the upper atmosphere, $\frac{N_t}{N_i}$, where $N_t$ is the total number of particles at timestep $t$, and $N_i$ is the initial total number of particles. For each molecule in each subplot, we plot the many flares case (green), which includes all flares in the lightcurve, and the few flares case (purple), which includes only flares with 25x quiescent flux in the lightcurve. The grey histograms are the overlapping areas between the many- and few flares cases.}
    \label{fig:manyfew}
\end{figure*}

The difference in the many- and fewer-flares simulations is illustrated in figure \ref{fig:manyfew} for the molecules SO$_2$, CH$_4$, CO$_2$, and COS. Each histogram represents the occurrence rate of the relative difference in abundance throughout the first $10^7$ seconds in the simulation. We interpolate the abundance of the molecules throughout time at 13000 set timesteps and determine the relative abundance at each of these timesteps for both datasets. Here, we define abundance as the total number of particles in the upper atmosphere (i.e., where most molecular changes take place depending on species, as indicated by the dashed orange lines in figure \ref{fig:manyfew}) by multiplying the number density at a certain pressure level by the volume of that layer. The green histogram in each subplot represents the occurrence rate of the simulation with all flares included.
In contrast, the purple histogram represents the occurrence rate of the simulation with only flares included that have a flux that is $>25\times$ the quiescent normalized flux. It can be seen from this figure that for all molecules, the green bar plots are higher for more significant relative differences. This indicates that while the large flares are the dominant ones that drive the change in the chemistry, minor flares also impact the results. The more minor flares likely build up a minor but shift in composition, which then translates into more drastic changes due to more prominent flares that have a more difficult time recovering afterward. However, the maximum relative difference for SO$_2$ and CH$_4$ remains the same, hinting that it does not accumulate more over time when more minor flares are included. For CO$_2$ and COS, this is not precisely the case. These molecules show a slight shift in the maximum relative difference in the total number of particles and go to more extreme levels. Although insignificant on§ these time scales, this indicates that adding more minor flares affects the accumulating effect for some species in the long term. 

\begin{figure*}
    \centering
    \includegraphics[width=1.0\linewidth]{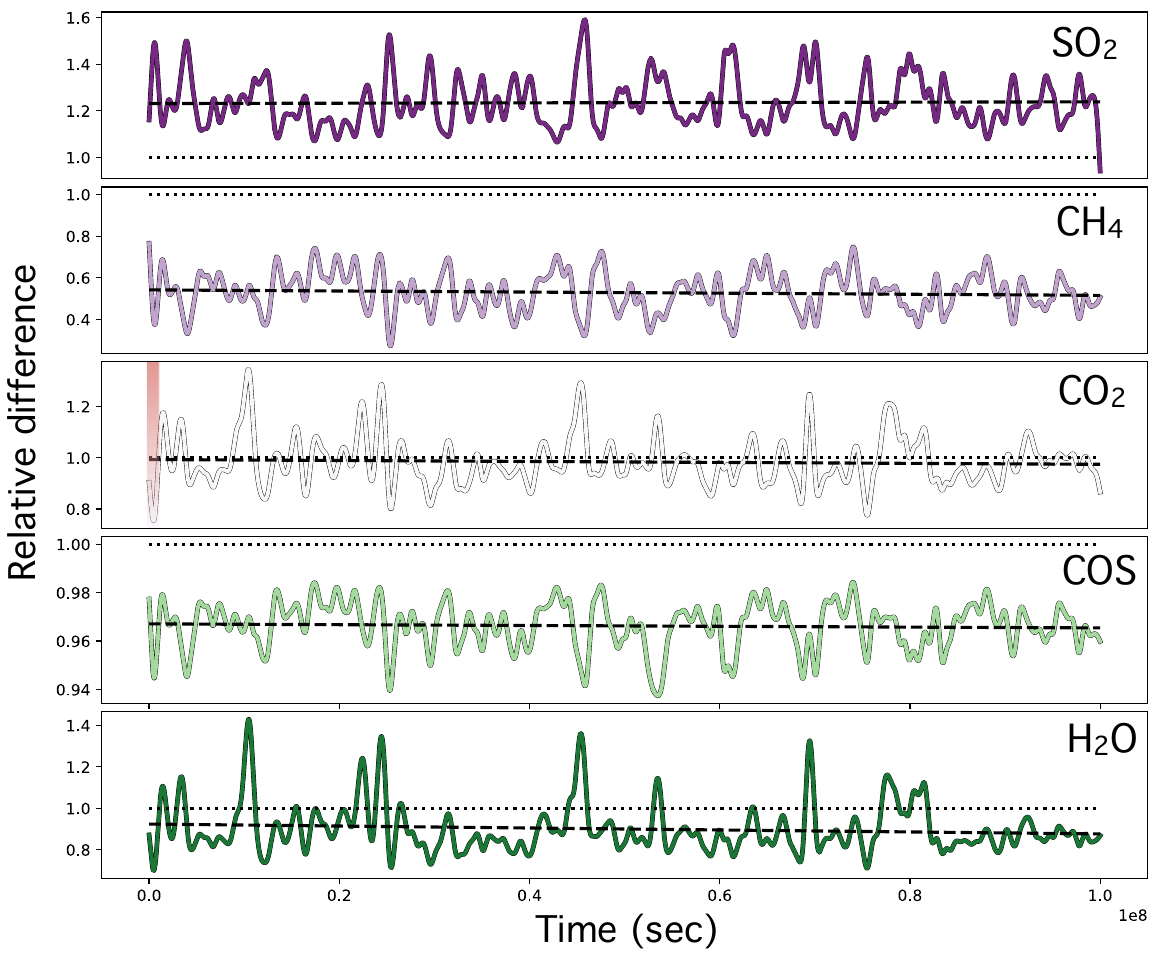}
    \caption{The variability (i.e., the relative difference of the abundance of key species at time $t$ and initial abundance) as a function of time. The abundance here is the total number of particles in the upper atmosphere. Each subplot represents the evolution of a different molecule. Note that these plots are smoothed with a Gaussian kernel to make the overall trend more visible. The horizontal dotted black lines represent the initial level, while the dashed black lines represent the linear fit of the relative abundance as a function of time. The red region in the subplot of CO$_2$ shows the integration time of L23, emphasizing the total integration time in this study.}
    \label{fig:longterm}
\end{figure*}

\subsubsection{Long-term accumulation effect}

L23 showed hints of a cumulative effect in the changing composition of atmospheres. In particular, they found that the relative abundances of CO$_2$ changed gradually over $\sim$11 days due to flaring events. In this study, we investigate this further by examining how the abundances change over a more extended period of $10^8$ seconds (i.e. $\sim$ 3 years). The relative difference in abundance, similar as described in section \ref{sec:morefewer}, as a function of time is shown in figure \ref{fig:longterm} for the molecules SO$_2$, CH$_4$, CO$_2$, COS, and H$_2$O. At first glance, these plots do not show a significant trend. The most prominent trend can be found in the H$_2$O abundance. After $\sim 3$ years into the simulation, the average H$_2$O abundance has decreased from 92\% of the initial abundance to 88\% of the initial abundance. Extrapolating this trend over time gives a decrease to 50\% of the initial abundance after $\sim$ 28 years. A similar but less significant trend can be seen for CH$_4$ where the abundance decreased by 2\% in $\sim 3$ years. To half this abundance by 50\%, an integration time of $\sim$ 32 years is needed. This extrapolation makes use of a linear trend, and is an oversimplification to the real scenario. However, it also gives us an extreme outline.

\textbf{SO$_2$}, \textbf{CH$_4$}, and \textbf{COS} show an elevated or lowered abundance over time. For SO$_2$, this is an elevated abundance level of $\sim$20\% higher than its original abundance, while for CH$_4$ and COS, this is a lowered abundance of $\sim$48\% and $\sim$3\%, respectively. 

\subsection{Transmission spectrum}

\begin{figure*}
    \centering
    \includegraphics[width=1.0\linewidth]{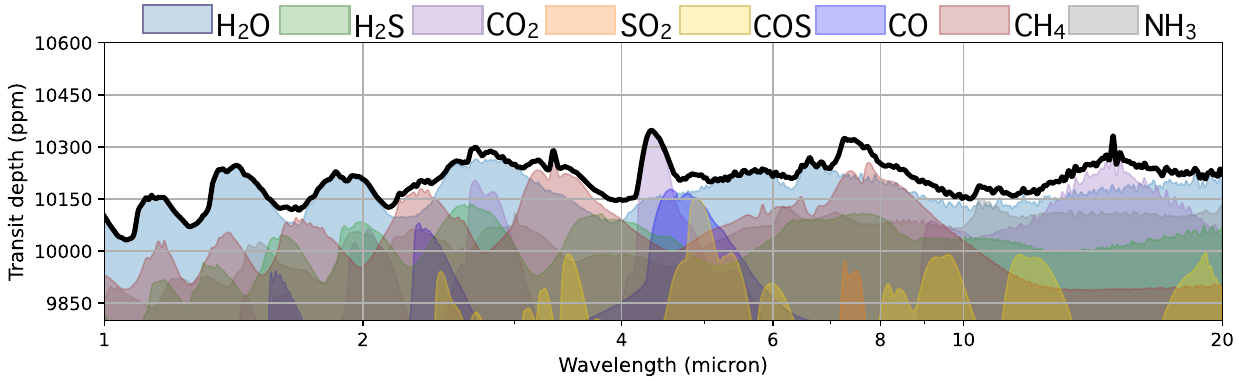}
    \caption{The quiescent transmission spectrum of a 1 Jupiter mass planet with a metallicity that is 10x solar elemental abundance.}
    \label{fig:quiescent_spectrum}
\end{figure*}

\subsubsection{Quiescent state}

The quiescent transmission spectrum, where no flare events are considered, is shown in figure \ref{fig:quiescent_spectrum}.  This spectrum highlights notable high-metallicity features. The molecule associated with atmospheric metallicity in the spectrum is \textbf{H$_2$O}, which exhibits distinct features across the 1 to 20 micron wavelength range. Other molecules of significance include \textbf{CO$_2$}, with features around 4.2 microns and 14 microns; \textbf{CO}, visible around 4.8 microns; and \textbf{SO$_2$}, showing between 7 - 8 microns, though notably absent near 4 microns. Additionally, the most prominent \textbf{COS} feature occurs around 5 microns but is challenging to distinguish from the CO feature at this wavelength. A prominent low-metallicity feature in the spectrum comes from \textbf{CH$_4$}, which appears between 2 - 4 microns and near 8 microns.  

\subsubsection{Extreme flare event}
\begin{figure*}
    \centering
    \includegraphics[width=1.0\linewidth]{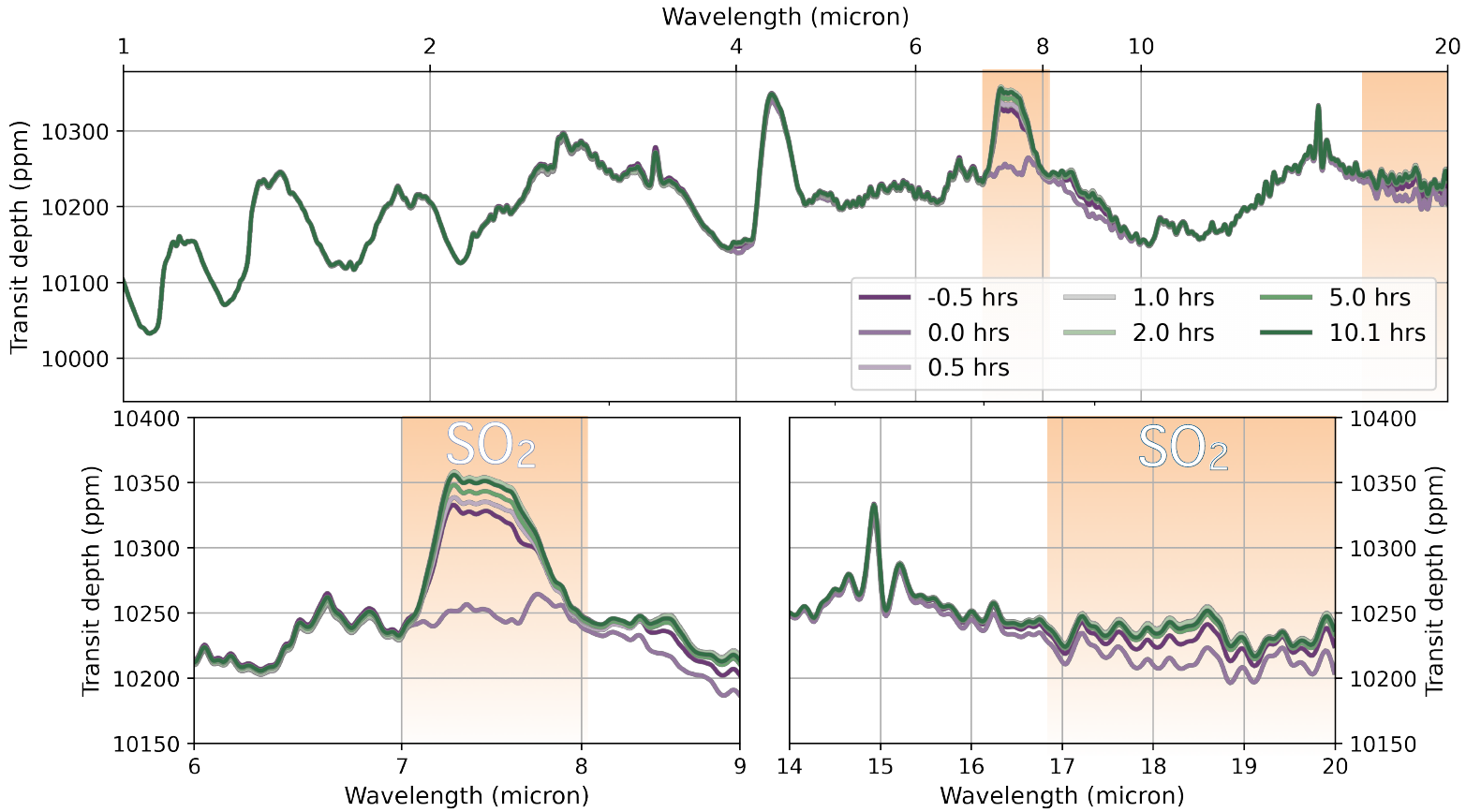}
    \caption{The change in transmission spectrum for an extreme flare event (see fig. \ref{fig:sulfursp_extreme_case}) in the IR-waveband (1 - 20 micron), as seen in the top panel. The spectrum is shown for several timesteps, starting at 30 minutes before the flare event and ending 10 hours after the event. The lower panels show a zoomed view of specific molecular features that show the most drastic changes within the spectrum, as indicated by the color patches.}
    \label{fig:spectrum_extreme}
\end{figure*}

The change in the transmission spectrum during an extreme flare event is shown in figure \ref{fig:spectrum_extreme}. The top panel shows that the most significant changes can be found in \textbf{SO$_2$} at 7 - 8 microns and 17 - 20 microns. The lower panels show a zoomed version of this molecule. The changing feature at longer wavelengths (17 - 20 microns) is less noticeable due to the more prominent water feature in that region (see figure \ref{fig:quiescent_spectrum}). The most considerable variability due to an extreme flare event can be seen in the 7-8 micron feature. At this wavelength range, the SO$_2$ feature disappears. This is in alignment with the extreme depletion of SO$_2$ around $10^{-3}$ bar $< P <$ $10^{-5}$ bar, as seen in figure \ref{fig:sulfursp_extreme_case}. 

The SO$_2$ feature overshoots the initial state and only converges $\sim 10$ hours after the flare event. This agrees with the abundance profiles, as seen in figure \ref{fig:sulfursp_extreme_case}. 
The large flare event briefly causes the SO$_2$ feature between 7-8 micron to disappear, with a difference of $\sim$ 75 ppm to what we expect to see at this wavelength. Briefly after, the overproduction over SO$_2$ causes the feature to return and even be enhanced compared to it's initial state with $\sim$ 25 ppm. 

\subsubsection{Variability}

\begin{figure*}
    \centering
    \includegraphics[width=1.0\linewidth]{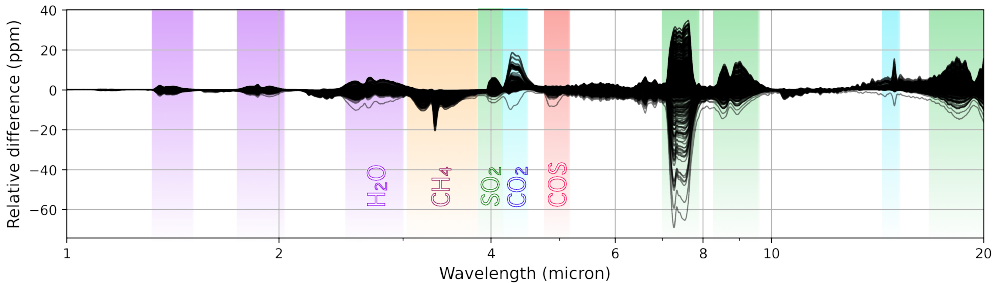}
    \caption{The relative difference (in ppm) between the spectrum at timestep $t$ and the quiescent state for 1000 random timesteps within the simulation. Each color patch represents the region of a certain molecule, as indicated in the figure.}
    \label{fig:variability}
\end{figure*}

The variability in the transmission spectrum of the planet's atmosphere is shown for 1000 random timesteps in the simulation in figure \ref{fig:variability}. Here, we see what the variability is with respect to a transmission spectrum that excludes stellar flaring. The most significant change occurs in the \textbf{SO$_2$} feature between 7-8 microns, with a maximum decrease of $\sim60$ ppm and a maximum increase of $\sim40$ ppm, and giving a total maximum variability of $\sim100$ ppm. The decrease in SO$_2$ features can be attributed to the photodissociation of this molecule due to the increased EUV irradiation during stellar flaring. However, the increase in SO$_2$ is more notable and somewhat more complex, as explained in section \ref{sec:extreme_flare_event_chem}. Other interesting variable features are the depletion of \textbf{CH$_4$}, between 3 - 4 micron. 

Figure \ref{fig:variability_SO2} shows how often we expect a specific spectral variability in the SO$_2$ feature at 7 - 8 microns over $10^7$ seconds divided by 13000 equally spaced time steps. As expected, it is relatively rare to observe a depletion of 60 ppm in the transmission spectrum due to flare events, which could potentially impact atmospheric retrievals. Similarly, it is uncommon to find the atmosphere in its quiescent state, with a relative difference of 0 ppm. In reality, we see a distribution with two distinct peaks. One can be attributed to flare events, and the other to recovery after these events. It is noticeable that the recovery peak is substantially more significant than the flare events peak. This is due to the long recovery rate compared to the photodissociation rate during flare events, which can also be seen in figures \ref{fig:sulfursp_extreme_case} and \ref{fig:spectrum_extreme}. Overall, we expect an elevated abundance in SO$_2$ due to flaring events, as shown in figure \ref{fig:longterm}.

\subsubsection{Impact on observations}

\begin{figure}
    \centering
    \includegraphics[width=1.0\linewidth]{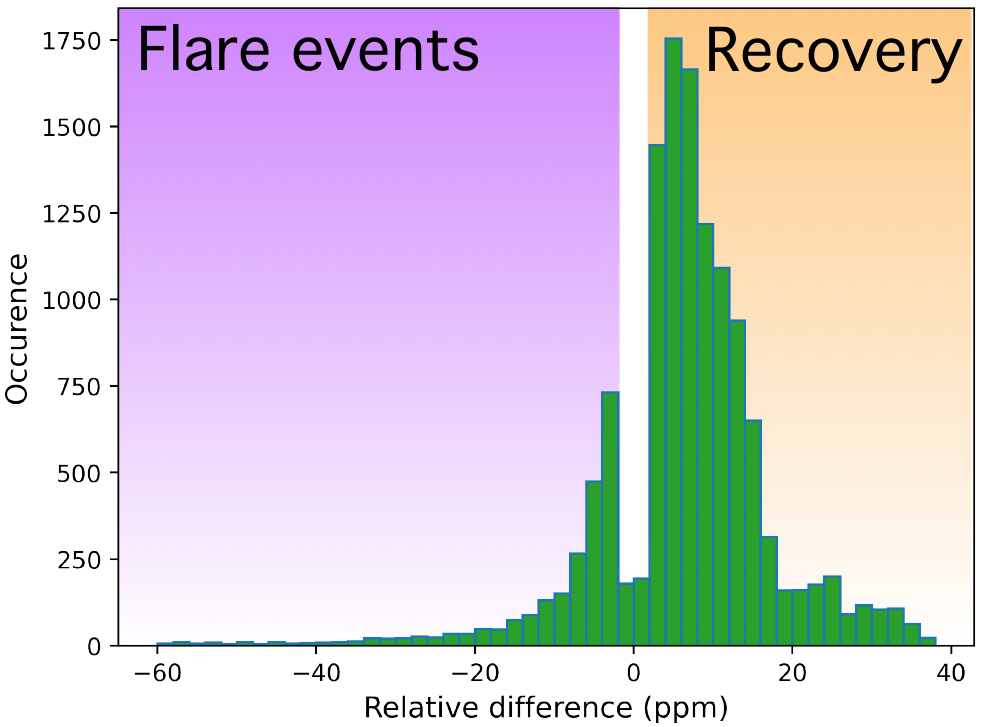}
    \caption{The occurrence of the variability in SO$_2$. The relative difference is the change in transmission spectrum in ppm at various time steps with respect to the quiescent transmission spectrum.}
    \label{fig:variability_SO2}
\end{figure}

In the previous sections it is shown that SO$_2$ exhibits the strongest variability of the species we consider, with a recovery timescale of $\sim$ 10 hours after a large flare. The transit of this system lasts only $\sim$ 1 hour \footnote{Calculated using $T \approx \frac{P}{\pi}\cdot \frac{R_{\star}}{a}$}, much shorter than the recovery time. Each transit spectrum is therefore a snapshot of the atmosphere at a random moment during the recovery, sampling the distribution shown in Figure \ref{fig:variability_SO2}. Whether this variability is detectable is a separate question: the change in transit depth between visits must exceed both the noise on the observations and the spectral changes caused by the star itself.

JWST observations of TOI-5205b, a close analogue of the system studied here (a $\sim$1 $M_{\mathrm{J}}$ planet around an M4 dwarf), illustrate how demanding this is \citep{Canas2026}. Over three transits within four days, the white-light transit depth was measured to a precision of $\sim$100 - 150 ppm, but only after fitting star spots whose configuration changed between every visit. More importantly, unocculted spots (the transit light source effect; \citeauthor{Rackham2018} \citeyear{Rackham2018}) altered the transmission spectrum by $\sim$5000 ppm in the optical and $\sim$1000 ppm at 5 $\mu$m, and \citet{Canas2026} found that the differences between their visits were caused by the star rather than the planet. Any search for planetary variability around an active M dwarf must therefore separate the planetary signal from an evolving stellar one that can be an order of magnitude larger.

For the system studied here, however, the amplitude of the modelled variability is small compared to what JWST can measure. The largest change occurs in the 7.3 $\mu$m SO$_2$ band and does not exceed $\sim$ 100 ppm, while the changes in the near-infrared, including the 4.0 $\mu$m band, remain below $\sim$20 ppm. Observing the 7 - 8 $\mu$m region requires MIRI, where the mid-infrared faintness of the M-dwarf host limits the precision per transit to several hundred ppm per spectral bin. Even when many transits are combined, JWST time-series observations reach a systematic noise floor of a few tens of ppm, which is comparable to the band-averaged variability signal itself. In addition, the star introduces its own spectral changes between visits at the $\sim$1000 ppm level even at 5 $\mu$m \citep{Canas2026}, an order of magnitude above the planetary signal modelled here. We therefore conclude that the SO$_2$ variability of this system is not detectable with JWST. A detection would require either a substantially brighter host star, a stronger driver of variability, or a future mid-infrared facility with a lower noise floor.

\section{Discussion}

\label{sec:DiscussionP4}

This study explored the three-year impact of recurrent stellar flares on the atmospheres of metal-rich, gas-rich exoplanets. Our findings indicate significant variability in the abundance profiles of various molecules, alongside variability in the transmission spectrum of key spectral features due to repeated flaring. The atmospheric composition and transmission spectra were rarely observed in a quiescent state. Additionally, we identified a declining trend in the abundances of water and methane, which, when extrapolated using a simple linear trend, suggested a halving of these molecules after approximately 29 to 31 years. However, the integration time was limited to only three years, making long-term projections from this dataset challenging. Extrapolating using a linear trend is therefore considered an extreme scenario.  Future research should investigate the extended impact of stellar flaring over longer periods.
To speed up these simulations, machine learning could be included to accelerate the chemical kinetics calculations. This method has proven to be effective for numerical photochemical kinetics calculations where static stellar spectra were included (\citeauthor{Hendrix2023} \citeyear{Hendrix2023}) and can be further explored to include a time-dependent stellar spectral component. Furthermore, long-term stellar activity must be considered, including short- and long-term variations, such as stellar evolution and potential cycles similar to the Sun's. This could offer a more comprehensive understanding of the flare-induced atmospheric changes over time.
Another caveat to this study is that the use of a 1D atmospheric model neglects horizontal transport processes that could influence the results. For instance, \citeauthor{Tsai2023_2DModel} \citeyear{Tsai2023_2DModel} demonstrated that horizontal mixing tends to produce more homogeneously distributed atmospheres. Incorporating such dynamics could dampen the variability in chemical abundances driven by stellar flaring, as perturbations would be redistributed more efficiently across the atmosphere.

In addition, we do not include photochemical haze in our models. Because small hydrocarbon
haze particles produce opacity that falls steeply toward longer wavelengths,
haze is expected to contribute little at the mid-infrared wavelengths where our
key features lie (in particular the 7--8\,$\mu$m SO$_2$ feature, \citeauthor{Kawashima2019} \citeyear{Kawashima2019}). However, a fully self-consistent haze
treatment is left to future work

Finally, we consider the observability of the modelled variability with current facilities. The strongest signal, the $\sim$  100 ppm change in the 7 - 8 $\mu$m SO$_2$ band, lies below the per-transit precision achievable with JWST/MIRI for a faint M-dwarf host, and repeated visits are ultimately limited by the systematic noise floor of JWST time-series observations. Moreover, recent JWST observations of TOI-5205b, a close analogue of the system studied here, showed that stellar activity alters the transmission spectrum at the $\sim$1000 ppm level even at 5 $\mu$m, with differences per visit attributable to evolving star spots rather than the planet \citep{Canas2026}. Any flare-driven planetary variability would therefore have to be disentangled from a stellar signal an order of magnitude larger, for example through a multi-epoch retrieval in which the stellar contamination is fitted separately for each visit \citep{Canas2026}.

\section{Conclusion}

\label{sec:ConclusionP4}

In this work, we modeled the impact of recurrent stellar flares on the atmospheres of metal-rich gaseous exoplanets for approximately three years to assess the long-term effects and potential variability in exoplanet atmospheres. By combining synthetic flare spectra, generated using a fiducial flare model (\citeauthor{Loyd2018a} \citeyear{Loyd2018a}), to a photochemical kinetics code, we tracked the evolution of atmospheric abundance profiles over time. These profiles were then forwarded to a radiative transfer code at various intervals to investigate how flare-induced changes affect spectral variability.  

Our simulations revealed persistent shifts in the abundances of key species, including CH$_4$, CO$_2$, and SO$_2$. A detailed analysis of a single extreme flare event showed an immediate and significant depletion of several molecular species in the upper atmosphere, with effects lasting approximately 30 minutes. This resulted in the temporary disappearance of certain spectral features, most notably the SO$_2$ feature at 7-8 microns, which experienced a significant shift of about 75 ppm. Post-flare, many molecular species did not return to their pre-flare (quiescent) state, leading to changes in their abundance profiles. This was further reflected in the time-dependent transmission spectrum, where the SO$_2$ feature converged after approximately 10 hours, showing variability of 100 ppm. 

Although the atmospheric composition largely returned to its quiescent state within 30 minutes after an extreme flare event, the probability of observing the transmission spectrum in this extreme state is low. Conversely, we also showed that the continuous variability introduced by recurrent flaring suggests that observing the atmosphere in its quiescent state is even less likely. Species such as CH$_4$, SO$_2$, CO$_2$, and COS showed the greatest spectral variability. SO$_2$ exhibited the most significant enhancement by a few factors, while CO$_2$ also displayed elevated levels. In contrast, CH$_4$ experienced a sustained depletion due to ongoing photodissociation in the upper atmosphere. This could explain depletion of CH$_4$ in real atmospheres when compared with their ideal, quiescent state. 

We also examined the cumulative long-term effects of recurrent flaring. Over time, the abundance of H$_2$O and CH$_4$ exhibited a decreasing trend, with a linear fit suggesting that these molecules would take approximately 28 to 31 years to reach half of their initial abundances. The linear trend in CO$_2$ found in L23 was not found in this study. This is most likely caused by the short integration time in L23. 

In conclusion, this study has shown the substantial long-term impacts of recurrent stellar flares on exoplanetary atmospheres, revealing both enduring and transient changes in atmospheric composition and spectral features. Direct detection of this variability remains out of reach with current instrumentation: the strongest signal, the $\sim$ 100 ppm change in the 7 - 8 $\mu$m SO$_2$ band, falls below the per-transit precision of JWST/MIRI for a faint M-dwarf host and is an order of magnitude smaller than the spectral changes introduced by stellar activity itself \citep{Canas2026}. The implication of our results is therefore not that the variability can be observed, but that it cannot be ignored: because the atmosphere is rarely in its quiescent state, any single observation samples a flare-perturbed composition, and abundances retrieved under the assumption of a quiescent atmosphere may be systematically biased. Flare activity should therefore be taken into account when characterizing the atmospheres of planets around active stars, even when the variability itself is undetectable.

\section*{Acknowledgements}
We acknowledge support from the European Research Council (ERC) under the European Union’s Horizon 2020 research and innovation programme (grant agreement no. 101088557, N-GINE).SMT is supported by the National Science and Technology Council (grants 114-2112-M-001-065-MY3) and Academia Sinica Career Development Award (AS-CDA-115-M03).
\section*{Data Availability}
The data presented in this article will be shared on reasonable request to the corresponding author. This work made use of publicly available, open-source codes for the photochemical and radiative transfer modeling. 


\bibliographystyle{mnras}
\bibliography{references} 




\appendix


\bsp	
\label{lastpage}
\end{document}